\documentclass[12pt, a4paper]{article}
\usepackage{graphicx}
\usepackage{amssymb}
\usepackage{stmaryrd}
\usepackage{enumitem}

\usepackage{longtable}
\usepackage[utf8]{inputenc}
\usepackage{booktabs} 
\usepackage{caption}
\usepackage{geometry}
\usepackage{tikz}
\usepackage{float}
\usepackage{url}
\usepackage{mathtools}
\usepackage{amsmath,amsfonts,amsthm}
\usepackage{mathrsfs}

\usepackage[utf8]{inputenc}
\usepackage{rotating}
\usepackage{multirow}
\usepackage{multirow}
\usepackage[table]{xcolor}
\newtheorem{remark}{Remark}

\usepackage{subfig}

\usepackage{ytableau}
\usepackage[colorlinks,
linkcolor=blue,
anchorcolor=blue,
citecolor=blue
]{hyperref}
\begin{document}
\begin{center}
{\large \bf Assessing Financial Statement Risks among $\mathrm{MCDM}$ Techniques}
\end{center}
\begin{center}
 Marwa Abdullah$^1$ \hspace{0.2 cm} Revzon Oksana Anatolyevna$^{2}$ \hspace{0.2 cm}  Duaa Abdullah$^{3}$\\[6pt]
 $^{1,2}$Department of Finance, State University of Management, 99 Ryazansky Prospekt, 109542, Moscow, Russia\\[6pt]
 $^{3}$Department of Discrete Mathematics, Moscow Institute of Physics and Technology, 141701, Moscow region, Russia\\[6pt]
Email: 	 $^{1}${\tt marwa1987abdullah2@gmail.com}, $^{2}$ {\tt revzonoks@yandex.ru},
 $^{3}${\tt duaa1992abdullah@gmail.com}
\end{center}
\noindent
\begin{abstract}
In this paper, to determine the financial risks faced by an industrial company, assessing the relative importance of these risks and identifying the years most exposed to financial risk using modern multi-criteria decision-making techniques. Applied to AL-Ahliah Vegetable Oil Company, the research utilizes the Analytical Hierarchy Process and Simple Additive Weighting to analyze financial ratios from 2008 to 2017.
\end{abstract}


\section{Introduction}
The process of \textit{identifying financial risks} represents an important part of the performance evaluation process in companies. 
To enable companies to form a clear picture of the \textit{financial risks} that may hinder their success, and to work on reducing those risks by making appropriate managerial decisions to confront them. 
It was necessary to highlight the process of identifying financial risks, focusing on the financial statements that represent the output of accounting work.

Since company management faces many complex problems, it needs fast and accurate methods to help solve these problems. 
The most prominent of these methods are multi-criteria decision-making methods, which are considered modern and simple methods that enable the company to divide the problem into simpler parts that facilitate its solution.

The use of \textit{multi criteria decision making} denote by $\mathrm{MCDM}$ methods in identifying the financial risks derived from the \textit{financial statements} enables determining the relative importance of those risks and identifying the riskiest among them.
This helps \textit{decision-makers} in companies to address them in a timely manner due to the impact of financial risks on the company's continuity and its ability to achieve its objectives.

Multiple categories, including market risk, credit risk, funding risk, liquidity risk, and cash flow risk introduced by Woods~\cite{Woods2008}. Market risk, credit risk, liquidity risk, operational risk, and other risks, including reputational risk and strategic risk introduced by B\"{o}blingen~\cite{Risks2008}. Based on the financial asset valuation model, which shows the relationship between the concept of these risks and the extent to which they can be eliminated, they have been classified into systematic risks and unsystematic risks by Zahra~\cite{Hamdani2012}.
They have been divided into three sections, mainly covering capital structure risk, liquidity risk, and long-term stability risk. This is based on the assumption that financial risks are those related to financing decisions in the company Btach~\cite{Btach2010}.

The research aims to identify the financial risks faced by the studied company derived from the financial statements. Use the Analytical Hierarchy Process ($AHP$) to determine the relative importance of the financial risks. Clarify how to integrate $AHP$ and the Simple Additive Weighting ($SAW$) method to identify the financial risks faced by the studied company. Identify the financial year most exposed to financial risks and the reasons for it.
\section{Statement Problem}
The importance of the research lies in the following aspects:
\begin{enumerate}
    \item Highlighting an important topic, which is the financial risks derived from the financial statements, as it is a topic that occupies the attention of scientific and practical circles due to the high financial risks in institutions and companies of different types.
    \item Revealing the financial risks that affect the future of the studied company through its financial statements.
    \item Determining the degree of relative importance of financial risks, i.e., converting these risks into numerical values that enable comparison between them.
    \item Helping the company identify the riskiest financial risks using modern methods, enabling decision-makers to prepare to face those risks.
\end{enumerate}
Due to the need of \textit{National Vegetable Oil Company} ($NVOC$) to know the financial risks it may face and the resulting future challenges, and because of the increasing importance of \textit{financial risks} for the company’s decision makers, which ensures the protection and continuity of the company, the problem of identifying the financial risks that the company may be exposed to has emerged. This can be expressed by answering the following questions:
\begin{enumerate}
    \item What are the financial risks that the company is exposed to, which can be inferred from the financial statements?
    \item To what extent does AHP contribute to determining the relative importance of financial risk?
    \item Is it possible to integrate the Analytical Hierarchy Process (AHP) and the Simple Additive Weighting (SAW) method in identifying the financial risks the company faces?
    \item Which financial years are more exposed to financial risks, and why?
\end{enumerate}

\section{Theoretical Framework of the Research}~\label{sec:Theoretical}

Financial risks are defined as the loss that may be exposed to due to uncertain changes. They are also defined as the company’s inability to meet its financial obligations, which results when the company relies on borrowing to finance its operations; these risks increase with the company's debts and its reliance on loans \cite{Bourbia2012}. In accounting thought, it is defined as ``the degree of uncertainty about future cash flows'' \cite{Suleiman2014}.

\subsection{Financial risks faced by companies}
Studying a company's financial statements is extremely important, as they provide a comprehensive picture of the company's activities, how it utilises its resources, its success in achieving its objectives (sales growth and profitability), its ability to continue (by generating cash flows), and its development (in terms of market share growth and return on equity). It also helps to analyse the company's liquidity, profitability and financing structure (Baljbali~\cite{Beljbaliya2010}).
The researcher noted that previous classifications of financial risks did not include a classification based on financial statements. 
Therefore, a financial risk model derived from financial statements was developed, which helps to study the company's financial position, profitability, and the risks involved in that profitability and liquidity, in order to arrive at a good assessment of the company's financial position~\cite{Beljbaliya2010}. 
There are several methods for studying risks derived from financial statements, and financial analysis is one of the most important financial management tools based on financial statements~\cite{Tawfiq1987}, which are provided by the accounting system as a primary source of information. These translate the company's various activities into a set of objective figures and tell us about the company's performance, problems and future, especially information that helps to identify and analyse financial risks.

Financial risks here refer to risks that can be derived from studying and analysing the risks of the company's financial position, and studying and analysing the risks of its profitability and cash flows, i.e. they include all of the following financial risks: risks derived from the financial position statement, risks derived from the income statement, and risks derived from the cash flow statement. Since financial ratios are an important tool of financial analysis, they will be used to express financial risks, as they facilitate comparison between companies and financial years and play a supporting role in decision-making. Each financial ratio reflects a weakness
or strength in the company's position~\cite{Din2016}. These risks are as follows:
\subsection{Risks derived from the financial position statement}
These include:
    \begin{enumerate}
        \item \textbf{Capital structure risk [CSR]:}    
    Focuses on the combination of equity and debt to finance the company~\cite{Alipour2015}. It reflects the internal sources of financing and part of the external sources of financing on which the company relies in the long term~\cite{Tawfiq1987}. The risk in the capital structure is represented by the debt ratio of this structure, and if the company is unable to pay the cost of that debt and repay the principal on its due dates, it poses a risk to the company. 
Several financial ratios have been used to express capital structure risk, namely among Table~\ref{tab01Financial} as follows:
\begin{table}[H]
\centering
\begin{tabular}{|l|l|}
\hline
Total Debt / Shareholders' Equity & Short-term Debt / Shareholders' Equity \\ \hline
Shareholders' Equity / Long-term Debt & Retained Earnings / Assets \\ \hline
Assets / Long-term Debt & Total Debt / Assets \\ \hline
Total Debt / Long-term Debt & Shareholders' Equity / Net Fixed Assets \\ \hline
Current Assets / Net Fixed Assets & Total Assets / Shareholders' Equity \\ \hline
Net Working Capital / Shareholders' Equity  & \\ 
\hline
\end{tabular}
\caption{Financial Ratios Reflecting the Risks of the Capital Structure.}~\label{tab01Financial}
\end{table}
\item \textbf{Liquidity Risk [LR]:} Liquidity means the possibility of converting the asset into cash at any time without incurring any loss, as the company's inability to pay its obligations when due due to the absence of cash carries a loss~\cite{ghafoud2016}, which necessitates studying the risks arising from it because it is extremely important to assess the financial situation, and can be expressed in the following ratios:
\begin{itemize}
    \item Turnover ratio 
 \item Fast liquidity ratio 
 \item Cash readiness ratio
\end{itemize}
\item \textbf{Income Risk [IR]:} The income statement reflects a summary of the results of the company's operations during the financial period prepared for it, through which it is possible to study the risks of the company's profitability, whether from its sales or its own funds, through the following set of financial ratios among Table~\ref{tab01assets}:
\begin{table}[H]
    \centering
    \begin{tabular}{|l|l|}
\hline
Net Profit Before Interest / net profit after interest & Total profit / sales \\ \hline
Net profit/ sales  & Profit before interest and tax / sales \\ \hline
Net profit+ interest / total assets & Net Profit / equity \\ \hline
    \end{tabular}
    \caption{Financial risk ratios derived from the income statement}
    \label{tab01assets}
\end{table}
\item \textbf{Cash Flow Risk [CFR]:} The flow list is of great importance in making financial decisions, and it is considered one of the most important financial statements that helps to identify the financial conditions of the company and its role in providing information that is not provided by both the statement of financial position and the income statement. It shows the monetary impact of all the company's operational, investment and financing activities, which helps to indicate the strengths and weaknesses of the company's performance~\cite{Amal2013}.  It provides information that can be relied upon to judge the liquidity and continuity of the company~\cite{khamosi2015}, as its analysis reveals risks represented by the following ratios among Table~\ref{tab03statement}:
\begin{table}[H]
    \centering
    \small
    \begin{tabular}{|l|}
\hline
   Net cash flows from operating activities / total cash flows from investing and financing activities \\ \hline Net cash flows from operating activities / sales \\ \hline
  Net cash flows from operating activities / capital expenditures (outflows from investing activities) \\ \hline  Net cash flows from operating activities / current liabilities \\ \hline
Net cash flows from operating activities/ net profit \\ \hline Net cash flows from operating activities / total assets \\ \hline
Net cash flows from operating activities / equity \\ \hline Net cash flows from operating activities / long-term debt \\ \hline
Operating cash inflows / initial cash requirements \\ \hline Net cash flows from operating activities / fixed assets \\ \hline
Net cash flows from operating activities /total debt \\ \hline Net operating cash flows / cash distributions \\ \hline
Net cash flows from operating activities / net cash flows from investing activities \\ \hline
Net cash flows from operating activities / net cash flows from financing activities\\ \hline
    \end{tabular}
    \caption{Financial risk ratios derived from the statement of cash flows.}
    \label{tab03statement}
\end{table}
    \end{enumerate}

%===========================================
\section{Analytical Hierarchy process [AHP]}
%===========================================
Due to the importance of the decision-making process, which represents the core of management work in a company, many methods have been developed to help solve administrative problems. Some depend on the personal judgment of the decision-maker, while others rely on quantitative methods that simplify complex problems, with the best being those that combine both to reach a sound decision~\cite{Sultan2015}. The Analytic Hierarchy Process is considered a mathematical method for solving complex problems involving multiple criteria, designed and developed by the scientist Saaty in 1980. It is also a quantitative approach used in many diverse fields and countries where its effectiveness in solving complex problems has been proven~\cite{Sumbung2014}.

It was defined by Smojver as a method for determining the relative importance of criteria and identifying the preferences for each alternative according to a measurement scale, through a set of pairwise comparisons, with the possibility of decomposing a criterion into a group of sub-criteria~\cite{Smojver2011}. This process involves decomposing the problem into its hierarchical components consisting of several levels. At the top of the hierarchy, the goal representing the problem is defined. At the second level, the criteria representing the secondary parts of the problem are identified and called \textit{main criteria}, which may branch out into more precise criteria called \textit{secondary criteria}, which themselves may branch into further sub-criteria. The base of the hierarchy represents the set of studied alternatives ( see Table~\ref{tab01scale}). Depending on the constructed model, an expert questionnaire is prepared and filled out by experts in the problem field, through pairwise comparison of the criteria and determining the degree of importance of each criterion relative to the higher-level criterion based on the following scale developed by Saaty~\cite{Saaty2008} as we show that in Table~\ref{tab01scale}:
\begin{table}[H]
\centering
\renewcommand{\arraystretch}{1.3}
\begin{tabular}{|c|l|p{7cm}|}
\hline
\textbf{Degree of importance} & \textbf{Definition}        & \textbf{Explanation} \\ \hline
1    & Equal importance         & Both criteria contribute equally to the achievement of the objective. \\ \hline
3    & Weak importance          & One criterion is slightly preferred over the other. \\ \hline
5    & Strong importance        & One criterion significantly outweighs the other in importance. \\ \hline
7    & Very strong importance   & One of the criteria is much more important than the other. \\ \hline
9    & Absolute importance      & One of the criteria is absolutely more important than the other. \\ \hline
2,4,6,8 &                       & Medium importance between the above values. \\ \hline
\end{tabular}
    \caption{The scale of the hierarchical analysis process~\cite{zbek2015}.}
    \label{tab01scale}
\end{table}
For example, we ask the following question to the expert in Table~\ref{tab01readiness}: determine the degree of importance of the main criteria as risks to which the company may be exposed
\begin{table}[H]
\centering
\tiny
\begin{tabular}{|c|lll|lllllllllll|lll|l|}
\hline
\multirow{2}{*}{Main Standards}                                   & \multicolumn{3}{l|}{Positive}                       & \multicolumn{11}{c|}{Equality}                                                                                                                                                                                                                              & \multicolumn{3}{l|}{Nigative}                       & \multicolumn{1}{c|}{\multirow{2}{*}{Main Standards}} \\ \cline{2-18}
                                                                  & \multicolumn{1}{l|}{9} & \multicolumn{1}{l|}{8} & 7 & \multicolumn{1}{l|}{6} & \multicolumn{1}{l|}{5} & \multicolumn{1}{l|}{4} & \multicolumn{1}{l|}{3} & \multicolumn{1}{l|}{2} & \multicolumn{1}{l|}{1} & \multicolumn{1}{l|}{2} & \multicolumn{1}{l|}{3} & \multicolumn{1}{l|}{4} & \multicolumn{1}{l|}{5} & 6 & \multicolumn{1}{l|}{7} & \multicolumn{1}{l|}{8} & 9 & \multicolumn{1}{c|}{}                                \\ \hline
\multirow{3}{*}{Risks from changing the main shape (Criterion 1)} & \multicolumn{1}{l|}{}  & \multicolumn{1}{l|}{}  &   & \multicolumn{1}{l|}{}  & \multicolumn{1}{l|}{}  & \multicolumn{1}{l|}{}  & \multicolumn{1}{l|}{}  & \multicolumn{1}{l|}{}  & \multicolumn{1}{l|}{}  & \multicolumn{1}{l|}{}  & \multicolumn{1}{l|}{}  & \multicolumn{1}{l|}{}  & \multicolumn{1}{l|}{}  &   & \multicolumn{1}{l|}{}  & \multicolumn{1}{l|}{}  &   & Criterion 2                                          \\ \cline{2-19} 
                                                                  & \multicolumn{1}{l|}{}  & \multicolumn{1}{l|}{}  &   & \multicolumn{1}{l|}{}  & \multicolumn{1}{l|}{}  & \multicolumn{1}{l|}{}  & \multicolumn{1}{l|}{}  & \multicolumn{1}{l|}{}  & \multicolumn{1}{l|}{}  & \multicolumn{1}{l|}{}  & \multicolumn{1}{l|}{}  & \multicolumn{1}{l|}{}  & \multicolumn{1}{l|}{}  &   & \multicolumn{1}{l|}{}  & \multicolumn{1}{l|}{}  &   & Criterion 3                                          \\ \cline{2-19} 
                                                                  & \multicolumn{1}{l|}{}  & \multicolumn{1}{l|}{}  &   & \multicolumn{1}{l|}{}  & \multicolumn{1}{l|}{}  & \multicolumn{1}{l|}{}  & \multicolumn{1}{l|}{}  & \multicolumn{1}{l|}{}  & \multicolumn{1}{l|}{}  & \multicolumn{1}{l|}{}  & \multicolumn{1}{l|}{}  & \multicolumn{1}{l|}{}  & \multicolumn{1}{l|}{}  &   & \multicolumn{1}{l|}{}  & \multicolumn{1}{l|}{}  &   & Criterion 4                                          \\ \hline
\multicolumn{1}{|l|}{\multirow{2}{*}{Criterion 2}}                & \multicolumn{1}{l|}{}  & \multicolumn{1}{l|}{}  &   & \multicolumn{1}{l|}{}  & \multicolumn{1}{l|}{}  & \multicolumn{1}{l|}{}  & \multicolumn{1}{l|}{}  & \multicolumn{1}{l|}{}  & \multicolumn{1}{l|}{}  & \multicolumn{1}{l|}{}  & \multicolumn{1}{l|}{}  & \multicolumn{1}{l|}{}  & \multicolumn{1}{l|}{}  &   & \multicolumn{1}{l|}{}  & \multicolumn{1}{l|}{}  &   & Criterion 3                                          \\ \cline{2-19} 
\multicolumn{1}{|l|}{}                                            & \multicolumn{1}{l|}{}  & \multicolumn{1}{l|}{}  &   & \multicolumn{1}{l|}{}  & \multicolumn{1}{l|}{}  & \multicolumn{1}{l|}{}  & \multicolumn{1}{l|}{}  & \multicolumn{1}{l|}{}  & \multicolumn{1}{l|}{}  & \multicolumn{1}{l|}{}  & \multicolumn{1}{l|}{}  & \multicolumn{1}{l|}{}  & \multicolumn{1}{l|}{}  &   & \multicolumn{1}{l|}{}  & \multicolumn{1}{l|}{}  &   & Criterion 4                                          \\ \hline
\multicolumn{1}{|l|}{Criterion 3}                                 & \multicolumn{1}{l|}{}  & \multicolumn{1}{l|}{}  &   & \multicolumn{1}{l|}{}  & \multicolumn{1}{l|}{}  & \multicolumn{1}{l|}{}  & \multicolumn{1}{l|}{}  & \multicolumn{1}{l|}{}  & \multicolumn{1}{l|}{}  & \multicolumn{1}{l|}{}  & \multicolumn{1}{l|}{}  & \multicolumn{1}{l|}{}  & \multicolumn{1}{l|}{}  &   & \multicolumn{1}{l|}{}  & \multicolumn{1}{l|}{}  &   & Criterion 4                                          \\ \hline
\end{tabular}
\caption{Part of the readiness assessment questionnaire concerning main criteria.}~\label{tab01readiness}
\end{table}

\section{Steps of the Analytic Hierarchy Process (AHP)}~\label{sec:Steps}

\begin{enumerate}
    \item After constructing the hierarchical model pertaining to the problem and conducting pairwise comparisons to determine the importance of each criterion as assessed by experts, we proceed to apply the following steps~\cite{Turan2016}:
\begin{enumerate}
  \item Building a pairwise comparison matrix for each criterion: This matrix consists of the main diagonal elements, each considered as the comparison of an element to itself, while the elements above the main diagonal are determined based on the expert’s estimates from the expert questionnaire. The elements below the main diagonal are calculated as follows:
  \begin{enumerate}
      \item  $A_{ij}$ represents the element in row $i$ and column $j$ in matrix $A$, and the column $j$ in the matrix.
    \[
    A_{ij} = \frac{1}{a_{ji}}
    \]
   \item Calculate the sum of each column in the pairwise comparison matrix.
  \end{enumerate}
    \item \textbf{Derive the normalized comparison matrix}: Each element of the matrix is divided by the sum of its column.
    \item \textbf{Calculate the relative weights for each row in the previous matrix}: Add the elements of each row, then divide by the number of elements in the row. The resulting vector represents the relative weights of each criterion, which determines the relative importance of each criterion.

    \item \textbf{Calculate the consistency ratio or consistency index}~\cite{Tataa2014}: To examine the degree of agreement among expert estimates. If the contradiction degree among the expert’s estimates does not exceed 10\%, the estimates are acceptable; otherwise, the expert’s estimates need to be reviewed. The following steps are followed:
    \begin{enumerate}
        \item \textbf{Analytic Hierarchy Process (AHP) Account - Maximum Eigenvalue ($\lambda_{\max}$):} The maximum eigenvalue ($\lambda_{\max}$) is calculated using the following formula (where the matrix is square of order $n \times n$):
\begin{equation}~\label{eqqn1}
\lambda_{\max} = \sum_{i=1}^{m} EV_i * S_i,
\end{equation}
where:
\begin{itemize}
    \item $EV_i$ is the eigenvector corresponding to the element $S_i$ (i.e., the weight of the criterion or alternative).
    \item $S_i$ is the sum of the elements in column $i$ of the pairwise comparison matrix.
\end{itemize}

Thus, $\lambda_{\max} \geq n$.

\item \textbf{Consistency Index (CI):} The consistency index is calculated using Saaty's method for the analytic hierarchy process ($\lambda_{\max}$) as follows ($n$ is the matrix order):
\begin{equation}~\label{eqqn2}
CI=\frac{\lambda_{\max}-n}{n-1}.
\end{equation}

\item  \textbf{Random Consistency Index (RI):}
It is a value obtained from the random index Table~\ref{tab01Donegan} developed by Saaty (from random judgments).
\begin{table}[H]
\centering
\begin{tabular}{|l|c|c|c|c|c|c|c|c|}
\hline
\textbf{N} & 1 & 2 & 3 & 4 & 5 & 6 & 7 & 8 \\
\hline
\textbf{RI} & 0.00 & 0.00 & 0.58 & 0.90 & 1.12 & 1.24 & 1.32 & 1.41 \\ \hline
\textbf{N}& 9 & 10 & 11 & 12 & 13 & 14 &15&\\ \hline
\textbf{RI} & 1.45 & 1.49 & 1.51 & 1.48 & 1.56 & 1.57 & 1.59 & \\ \hline 
\end{tabular}
\caption{Random Consistency Index (RI) values~\cite{Donegan1991}.}~\label{tab01Donegan}
\end{table}
\item \textbf{Consistency Ratio (CR):} The consistency ratio is calculated as follows:
\[
CR = \frac{CI}{RI} < 0.10
\]
The result is considered acceptable if the consistency ratio is less than 0.10. If the consistency ratio is 0.10 or greater, the judgments must be revised and the pairwise comparison matrix re-evaluated.
\end{enumerate}
\end{enumerate}
\end{enumerate}
\begin{remark}
The consistency ratio is used to assess the reliability of the decision-maker's judgments. If the value is less than 0.10, the judgments are considered consistent and acceptable. If the value is 0.10 or higher, the decision-maker must review and revise the pairwise comparison matrix to improve consistency.
\end{remark}

\section{Simple Additive Weighting (SAW) method}
This is one of the most widely used multi-criteria decision-making methods and the most commonly used in evaluating alternatives. This method relies on weighted aggregation, where the evaluation score for each alternative is calculated by multiplying the measured value for each alternative by the relative weights of the criteria, which are determined based on the experience of decision-makers~\cite{Shin2013}.

\section{Steps of the simple weighted method}
This method refer to~\cite{Sultan2015}. Thus, we have 
\begin{enumerate}
    \item \textbf{Building the Pairwise Comparison Matrix $D = [X_{ij}]$ of order $m \times m$:}
It is constructed based on the relative importance of each criterion $C_i$ over criterion $C_j$, where $i, j = 1, 2, \ldots, m$, using the following matrix $D$ of order $m \times m$:
\[
D = 
\begin{bmatrix}
A_1 \\
A_2 \\
\vdots \\
A_m
\end{bmatrix}
\begin{bmatrix}
x_{11} & x_{12} & \cdots & x_{1m} \\
x_{21} & x_{22} & \cdots & x_{2m} \\
\vdots & \vdots & \ddots & \vdots \\
x_{m1} & x_{n2} & \cdots & x_{mm}
\end{bmatrix}_{m \times m}
\]
where $C_i$ is criteria for $i\in \mathbb{N}$, $A_{ij}$ is the value of replacement, and $X_{ij}$ the diagonal elements $X_{ij} = 1$ for all $i$.

\item \textbf{Converting the Pairwise Comparison Matrix into a Normalized Matrix (Max/Min):}

This is done by dividing each element in the column by the largest element in that column (if the criterion is of benefit type), or by the smallest element (if the criterion is of cost type), using the following:
\[
r_{ij} = 
\begin{cases} 
\dfrac{x_j^- - x_j}{x_j^+ - x_j^-} & ; \quad j \in \Omega_{\max} \\[10pt]
\dfrac{x_j - x_j^-}{x_j^+ - x_j^-} & ; \quad j \in \Omega_{\min},
\end{cases}
\]
Where:
\begin{itemize}
    \item $x_j^+$: the maximum value of criterion $j$
    \item $x_j^-$: the minimum value of criterion $x_j$
    \item $\Omega_{\max}$: set of benefit criteria
    \item $\Omega_{\min}$: set of cost criteria
\end{itemize}

\item \textbf{Building the Weighted Normalized Matrix $V$:}

This is done by multiplying the normalized matrix $V = [\tilde{x}_{ij}]$ by the vector of relative weights of the criteria $W = [w_i]$:

\[
V = [v_{ij}]_{m \times m} = 
\begin{bmatrix}
w_{11} & w_{12} & \cdots & w_{1m} \\
w_{21} & w_{22} & \cdots & w_{2m} \\
\vdots & \vdots & \ddots & \vdots \\
w_{m1} & w_{m2} & \cdots & w_{mm}
\end{bmatrix}
\]

\item \textbf{Calculate the performance vector of alternatives using the weighted normalized matrix $V$:}

\[
V = \sum_{j=1}^{m} w_jr_{ij}.
\]

\item \textbf{Rank the alternatives:}

\[
A_{ij}=\dfrac{V_{ij}}{\sum_{j=1}^{m}V_{ij}}.
\]

\item \textbf{Determine the best alternative.}
\end{enumerate}

\section{Previous Studies}

\begin{itemize}
\item \textbf{Study (Btach, 2010):} Titled \textit{``Identifying Financial Risks Based on Balance Sheet Information''}~\cite{Btach2010}. This study aims to define financial risks, their components, and factors derived from the information provided in the balance sheet. Financial ratios were used for 100 Polish companies over a period of 10 years (2000--2009). The study concluded that there are \textit{three components of financial risk} presented by balance sheet information: Capital structure risk, Liquidity risk and Long-term borrowing risk. These can be considered a foundation that companies use in financial planning and forecasting financial risks.

\item \textbf{Study (AbdelHamid R., Zain E., 2012):} Titled \textit{``Performance Evaluation Using Decision Support Systems''}~\cite{Abdelhamid2012}. The study presented a performance evaluation model based on decision support systems using financial ratios and \textit{three multi-criteria decision-making methods (AHP, SAW, TOPSIS)} applied to 8 public and private Egyptian pharmaceutical companies. Seven financial ratios were used as evaluation criteria for the period 2004--2011. The three methods were compared to determine the best-performing company.  The study concluded that \textit{SAW and TOPSIS gave the same company ranking when relying on AHP}, but produced \textit{different rankings when based on expert opinions}.

    \item \textbf{Study (Taataa, Samakiyeh, 2014):} Titled \textit{``Risk Management in Textile Industry Companies in Aleppo Using the Analytic Hierarchy Process (AHP)''}~\cite{Tataa2014}. This study aimed to identify the nature of risks facing projects, determine the most significant potential risks, and find the best strategies to address them using the \textit{Analytic Hierarchy Process (AHP)}. The researcher adopted a descriptive-analytical approach. A questionnaire was completed by \textbf{5 experts and decision-makers} at \textit{Hi-Tex Textile Industries Company} in Aleppo. The study concluded that \textit{technical risks} are the most dangerous factor for Hi-Tex, and the \textit{best strategy to address them is avoidance}.

    \item \textbf{Study (Krivka, Stonkute, 2015):} Titled \textit{``Comparative Analysis of Financial Position and Performance in Construction Companies''}~\cite{Krivka2015}. The study aimed to conduct a comparative analysis of the financial position and performance of two construction companies based on financial criteria. The \textit{Simple Additive Weighting (SAW)} method was applied using \textit{four groups of financial ratios}: Profitability, Liquidity, Solvency and Asset turnover. 
These were used as evaluation criteria. Weights were estimated by \textit{7 specialized experts} over three years (2011--2013) for two companies treated as evaluation alternatives. The study concluded that \textit{Merko Ehitus outperformed Nordicon} in financial position and performance, making it the \textit{better alternative}.

 \item \textbf{Study (Ghazani, Alishah, 2017):} Titled \textit{``Ranking Companies Based on Financial Indicators and Multi-Criteria Decision-Making Techniques -- A Case Study in Subsidiaries of the National Petrochemical Company in Iran''}~\cite{Ghazani2017}. The study aimed to rank subsidiary companies based on financial indicators and multi-criteria decision-making techniques to form a rational basis for evaluating their financial performance.  A comparison was made between \textit{AHP, SAW, TOPSIS, and Borda} methods. The descriptive-analytical and inductive approaches were used. A questionnaire was designed with \textit{19 key financial indicators}, and their relative importance was determined by a group of lecturers, PhD, and master’s students in accounting and financial management for the period \textit{2003--2008}.  The study found that the techniques used assigned \textit{different weights to financial indicators}, and all contributed to decision-making. These weights were \textit{close to the actual reality of the companies}.
\item \textbf{Study (Saarman, 2017):}
\textbf{Titled:} ``\textit{Financial Analysis of Performance and Risk -- Case of Svenska Handelsbanken AB}''~\cite{Saarman017}. This study aimed to examine the evolution and determine the relationship between performance and risk, as well as to identify the impact of the 2007 financial crisis on the development of financial performance. A \textit{mixed-method research approach} was employed for data collection and analysis. The balance sheet and income statement of the studied company were analyzed for the period \textit{2000--2016} using \textit{financial ratios for performance} (liquidity ratios, leverage ratios, borrowing ratios) and \textit{risk} (beta model, Value at Risk, Capital Asset Pricing Model). The study concluded that:
\begin{itemize}[leftmargin=*]
    \item The company’s \textit{performance was good} throughout the studied period, with \textit{stable liquidity} except for sudden spikes in values during a few years.
    \item The company was \textit{capable of meeting its short- and long-term obligations}.
    \item This was accompanied by an \textit{increase in the total debt ratio}, particularly in \textit{2006}.
    \item \textit{Funding was sound and diversified}.
    \item \textit{Profitability was in a good position}, though it \textit{declined sharply during the financial crisis of 2008--2009}.
    \item The study showed that the company had \textit{strong risk management} during and after the financial crisis.
    \item The \textit{financial crisis had no statistically significant impact} on performance or increase in risk.
\end{itemize}
\end{itemize}

\paragraph{What Distinguishes the Research from Previous Studies:}
\begin{itemize}
    \item Developing a new classification of financial risks based on the three financial statements: the statement of financial position, the income statement, and the statement of cash flows.
    \item Enriching Arabic studies with a study that relies on modern quantitative methods under multiple criteria, where the number of criteria exceeds twenty.
    \item Using actual historical values derived from the financial statements of the studied company to calculate financial ratios instead of expert estimates for these ratios.
    \item Integrating AHP and SAW to identify the financial risks facing the company.
    \item The studied alternatives are financial years, not companies.
\end{itemize}

\section{Case Study}

\subsection{Overview of Al-Ahlia Vegetable Oils Company}
Al-Ahlia Vegetable Oils Company is a public joint-stock company established in 1995. It is one of the largest oil companies in Syria, with a capital of 1,500 million SYP (equivalent to 30 million USD), fully self-financed. It has 3,282 shareholders. Its products are distinguished by high quality and compliance with international standards. The company produces vegetable ghee, vegetable oil, cottonseed hulls, sunflower meal, soapstock for soap manufacturing, and others.

Al-Ahlia Vegetable Oils Company represents the entire industrial sector listed on the Damascus Securities Exchange, thus holding a significant position in the Damascus financial market. Given the company's need to identify potential financial risks and work to avoid them—ensuring its continuity and disclosing such risks in its financial statements—a expert questionnaire was completed by the five financial experts in the company who play a key role in decision-making. They are:

\begin{table}[H]
\centering
\small
\begin{tabular}{|l|c|l|l|}
\hline
\textbf{Academic degree} & \textbf{Experience} & \textbf{Job Title} & \textbf{Name} \\
\hline
Diploma Economic & 14 & Direct sales & First alternative \\ \hline
Diploma Economic & 6 & Islamic banks & Second alternative \\ \hline
Diploma Economic & 6 & Banks and the stock market & Third alternative \\ \hline
Master Economic / financing & 8 & Banks and financing institutions & Fourth alternative \\ \hline
Diploma Economic & 9 & Private and public offering & Fifth alternative \\
\hline
\end{tabular}
\caption{Experience Financial in $NVOC$.}~\label{tabmarn1}
\end{table}

The financial statements of the studied company were also analyzed using financial ratios for the years 2008--2017.

\subsection{Research Methodology}
The researcher adopted the inductive approach by studying the financial statements to identify the financial risks faced by the studied industrial company, then building a model to assess the company’s financial risks that helps determine the relative importance of each risk. The financial years from 2008 to 2017 were then evaluated through the integration of the Analytic Hierarchy Process (AHP) and the Simple Additive Weighting (SAW) method.

\subsection{Data Collection Method}
The researcher prepared an expert questionnaire after gathering information on potential financial risks that an industrial company might face. The questionnaire was sent to the financial experts at Al-Ahlia Vegetable Oils Company. Additionally, the company’s financial statements published on the Damascus Securities Exchange website were used.

\subsection{Data Analysis Method}
The expert questionnaire was analyzed using Microsoft Excel by converting it into a set of mathematical matrices required for the Analytic Hierarchy Process (AHP). The researcher also analyzed the financial statements of the studied company over ten years (2008--2017) using financial ratios, which were then used to integrate AHP and SAW.

\subsection{Identifying Risks Derived from Financial Statements through Integration of AHP and SAW}
AHP will be used to evaluate the criteria, followed by SAW to evaluate the alternatives, with integration between the two methods.

To demonstrate the importance of AHP in determining the relative importance of criteria, its previously mentioned steps were applied to the expert questionnaires filled out by the financial experts at Al-Ahlia Vegetable Oils Company listed on the Damascus Securities Exchange. These experts have experience ranging from 6 to 14 years in financial management.
\begin{itemize}
    \item The hierarchical model was first constructed to suit the research problem, as shown in the following Figure~\ref{fig001}:
\end{itemize}
\begin{figure}[H]
    \centering
    \includegraphics[width=0.8\linewidth]{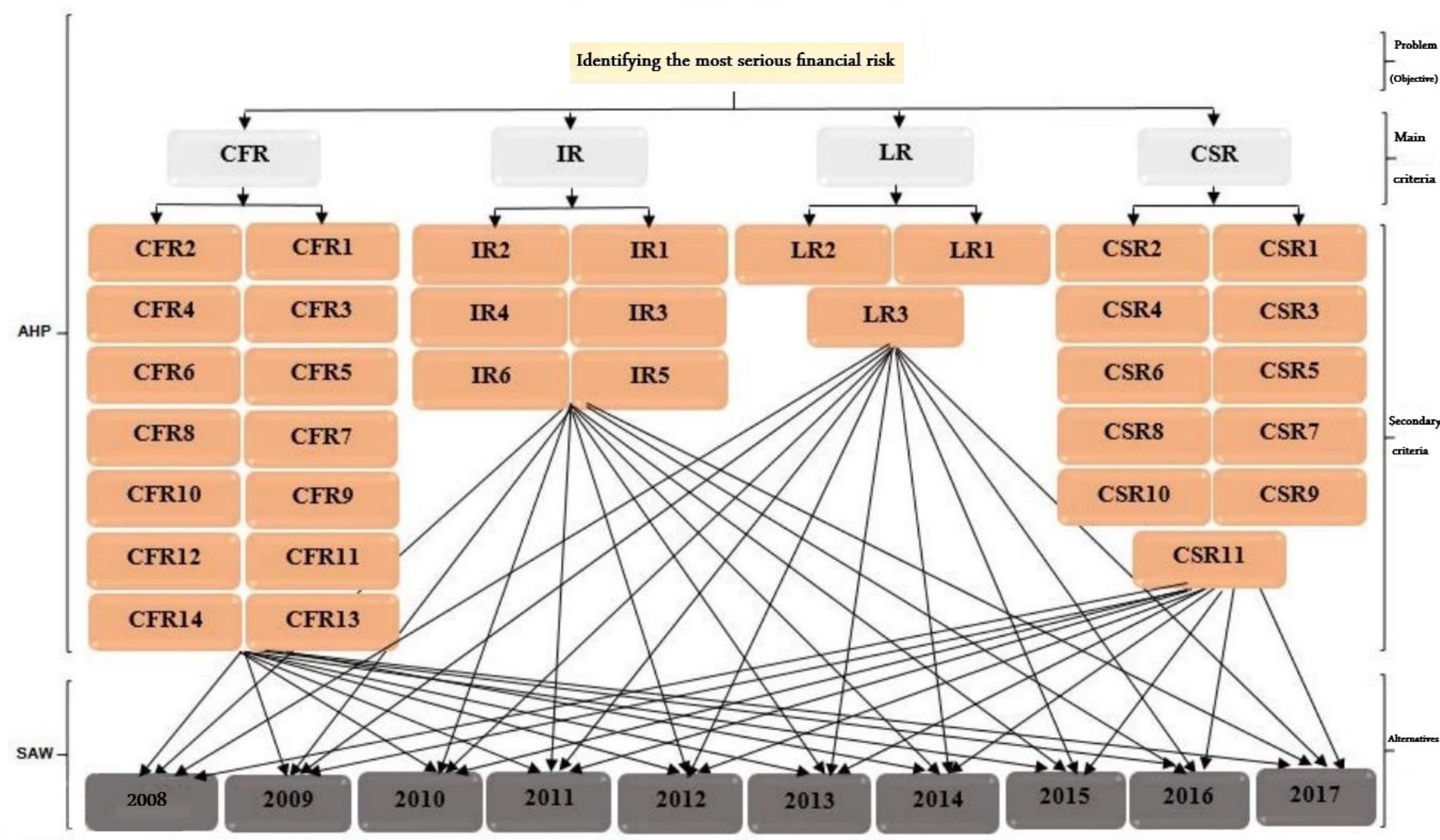}
    \caption{Hierarchical Model for Financial Risk Assessment.}
    \label{fig001}
\end{figure}
The following table shows the symbols used in the form Table~\ref{tab005Standardsr}:

\begin{table}[H]
\centering
\tiny
\begin{tabular}{|l|l|l|l|}
\hline
Symbol & Secondary                                                                                            & Symbol                                     & Primary                                   \\ \hline
CSR1   & Total debt / equity                                                                                  & \multicolumn{1}{c|}{\multirow{11}{*}{CSR}} & \multirow{11}{*}{Capital  structure Risk} \\ \cline{1-2}
CSR2   & Short-term debt / equity                                                                             & \multicolumn{1}{c|}{}                      &                                           \\ \cline{1-2}
CSR3   & Long-term debt / equity                                                                              & \multicolumn{1}{c|}{}                      &                                           \\ \cline{1-2}
CSR4   & Retained earnings / assets                                                                           & \multicolumn{1}{c|}{}                      &                                           \\ \cline{1-2}
CSR5   & Long-term debt / assets                                                                              & \multicolumn{1}{c|}{}                      &                                           \\ \cline{1-2}
CSR6   & Total debt / assets                                                                                  & \multicolumn{1}{c|}{}                      &                                           \\ \cline{1-2}
CSR7   & Long-term debt / total debt                                                                          & \multicolumn{1}{c|}{}                      &                                           \\ \cline{1-2}
CSR8   & Equity / net fixed assets                                                                            & \multicolumn{1}{c|}{}                      &                                           \\ \cline{1-2}
CSR9   & Invested funds / net fixed assets                                                                    & \multicolumn{1}{c|}{}                      &                                           \\ \cline{1-2}
CSR10  & Total assets / equity                                                                                & \multicolumn{1}{c|}{}                      &                                           \\ \cline{1-2}
CSR11  & Net working capital / equity                                                                         & \multicolumn{1}{c|}{}                      &                                           \\ \hline
LR1    & Turnover ratio                                                                                       & \multirow{3}{*}{LR}                        & \multirow{3}{*}{Liquidity Risk}           \\ \cline{1-2}
LR2    & Fast liquidity ratio                                                                                 &                                            &                                           \\ \cline{1-2}
LR3    & Cash readiness ratio                                                                                 &                                            &                                           \\ \hline
IR1    & Net Profit Before Interest / net profit after interest                                               & \multirow{6}{*}{IR}                        & \multirow{6}{*}{Income Risk}              \\ \cline{1-2}
IR2    & Total profit / sales                                                                                 &                                            &                                           \\ \cline{1-2}
IR3    & Net profit / sales                                                                                   &                                            &                                           \\ \cline{1-2}
IR4    & Net profit, net of interest and tax / sales                                                          &                                            &                                           \\ \cline{1-2}
IR5    & Net profit + interest / total assets                                                                 &                                            &                                           \\ \cline{1-2}
IR6    & Net Profit / equity                                                                                  &                                            &                                           \\ \hline
CFR1   & Net cash flows from operating activities / total cash flows from investing and financing activities  & \multirow{14}{*}{CFR}                      & \multirow{14}{*}{Cash Flow Risk}          \\ \cline{1-2}
CFR2   & Net cash flows from operating activities / income (sales)                                            &                                            &                                           \\ \cline{1-2}
CFR3   & Net cash flows from operating activities / capital expenditures (outflows from investing activities) &                                            &                                           \\ \cline{1-2}
CFR4   & Net cash flows from operating activities / current liabilities                                       &                                            &                                           \\ \cline{1-2}
CFR5   & Net cash flows from operating activities / net profit                                                &                                            &                                           \\ \cline{1-2}
CFR6   & Net cash flows from operating activities / total assets                                              &                                            &                                           \\ \cline{1-2}
CFR7   & Net cash flows from operating activities / equity                                                    &                                            &                                           \\ \cline{1-2}
CFR8   & Net cash flows from operating activities / Long-term debt                                            &                                            &                                           \\ \cline{1-2}
CFR9   & Operating cash inflows / initial cash requirements                                                   &                                            &                                           \\ \cline{1-2}
CFR10  & Net cash flows from operating activities / fixed assets                                              &                                            &                                           \\ \cline{1-2}
CFR11  & Net cash flows from operating activities / total debt                                                &                                            &                                           \\ \cline{1-2}
CFR12  & Net operating cash flows / cash distributions                                                        &                                            &                                           \\ \cline{1-2}
CFR13  & Net cash flows from operating activities / net flow from investing activities                        &                                            &                                           \\ \cline{1-2}
CFR14  & Net cash flows from operating activities / net cash flows from financing activities                  &                                            &                                           \\ \hline
\end{tabular}
\caption{Standards and their symbols.}
    \label{tab005Standardsr}
\end{table}
\begin{itemize}
\item  Then the experts ' data was unloaded using Microsoft Excel according to the steps of the hierarchical analysis process, and showed their application to the following five expert questionnaires:
Through Table~\ref{avertab001}, we denote \textit{Average weights of secondary criteria for the target (2)* (1)} by $AWS_1$, Average weights of secondary standards (2) by $AWS_2$  and average weights of the main criteria for the objective(1) by $AWS_3$
\begin{table}[H]
\begin{tabular}{|l|c|c|c|c|c|c|}
\hline
Rank                      & $AWS_1$ & SubRank                   & $AWS_2$ & Secondary criteria & $AWS_3$ & main criteria          \\ \hline
32                        & 0.0104                                                        & 9                         & 0.071225674                                & CSR1               &                                                           &                        \\ \cline{1-5}
28                        & 0.0119                                                        & 5                         & 0.081739877                                & CSR2               &                                                           &                        \\ \cline{1-5}
27                        & 0.0120                                                        & 4                         & 0.082536235                                & CSR3               &                                                           &                        \\ \cline{1-5}
34                        & 0.0099                                                        & 11                        & 0.067770123                                & CSR4               &                                                           &                        \\ \cline{1-5}
31                        & 0.0109                                                        & 8                         & 0.074968651                                & CSR5               &                                                           &                        \\ \cline{1-5}
33                        & 0.0101                                                        & 10                        & 0.069051707                                & CSR6               &                                                           &                        \\ \cline{1-5}
30                        & 0.0113                                                        & 7                         & 0.07789214                                 & CSR7               &                                                           &                        \\ \cline{1-5}
23                        & 0.0164                                                        & 2                         & 0.112913395                                & CSR8               &                                                           &                        \\ \cline{1-5}
26                        & 0.0125                                                        & 3                         & 0.086057521                                & CSR9               &                                                           &                        \\ \cline{1-5}
29                        & 0.0118                                                        & 6                         & 0.081255025                                & CSR10              &                                                           &                        \\ \cline{1-5}
17                        & 0.0283                                                        & \cellcolor{cyan}{1} & 0.194589852                                & CSR11              &                                                           &                        \\ \cline{1-5}
                          & 0.1456                                                        &                           & 1                                          & \textbf{SUM}       & \multirow{-12}{*}{0.14564}                                & \multirow{-12}{*}{CSR} \\ \hline
24                        & 0.0163                                                        & 3                         & 0.066884438                                & LR1                &                                                           &                        \\ \cline{1-5}
\cellcolor{red}{2} & 0.0944                                                        & 2                         & 0.387641292                                & LR2                &                                                           &                        \\ \cline{1-5}
\cellcolor{red}{1} & 0.1329                                                        & \textcolor{cyan}{1} & 0.54547427                                 & LR3                & \multirow{-3}{*}{0.24362}                                 & \multirow{-3}{*}{LR}   \\ \hline
                          & 0.2436                                                        &                           & 1                                          & \textbf{SUM}       &                                                           &                        \\ \cline{1-5}
22                        & 0.0196                                                        & 5                         & 0.129457616                                & IR1                &                                                           &                        \\ \cline{1-5}
25                        & 0.0156                                                        & 6                         & 0.102811315                                & IR2                &                                                           &                        \\ \cline{1-5}
20                        & 0.0232                                                        & 3                         & 0.152794825                                & IR3                &                                                           &                        \\ \cline{1-5}
19                        & 0.0251                                                        & 2                         & 0.165883471                                & IR4                &                                                           &                        \\ \cline{1-5}
21                        & 0.0231                                                        & 4                         & 0.15242835                                 & IR5                &                                                           &                        \\ \cline{1-5}
\cellcolor{red}{3} & 0.0450                                                        & \cellcolor{cyan}{1} & 0.296624623                                & IR6                &                                                           &                        \\ \cline{1-5}
                          & 0.1516                                                        &    & 1                                          & \textbf{SUM}       & \multirow{-8}{*}{0.15155}                                 & \multirow{-8}{*}{IR}   \\ \hline
12                        & 0.0313                                                        & 9                         & 0.068239078                                & CFR1               &                                                           &                        \\ \cline{1-5}
18                        & 0.0266                                                        & 14                        & 0.058013941                                & CFR2               &                                                           &                        \\ \cline{1-5}
7                         & 0.0343                                                        & 4                         & 0.074674756                                & CFR3               &                                                           &                        \\ \cline{1-5}
15                        & 0.0302                                                        & 12                        & 0.065822012                                & CFR4               &                                                           &                        \\ \cline{1-5}
\cellcolor{red}{4} & 0.0426                                                        &\cellcolor{cyan}{1} & 0.092697016                                & CFR5               &                                                           &                        \\ \cline{1-5}
16                        & 0.0283                                                        & 13                        & 0.061719932                                & CFR6               &                                                           &                        \\ \cline{1-5}
8                         & 0.0334                                                        & 5                         & 0.072842428                                & CFR7               &                                                           &                        \\ \cline{1-5}
11                        & 0.0318                                                        & 8                         & 0.069263673                                & CFR8               &                                                           &                        \\ \cline{1-5}
6                         & 0.0370                                                        & 3                         & 0.080643288                                & CFR9               &                                                           &                        \\ \cline{1-5}
13                        & 0.0307                                                        & 10                        & 0.066784395                                & CFR10              &                                                           &                        \\ \cline{1-5}
5                         & 0.0377                                                        & 2                         & 0.082128067                                & CFR11              &                                                           &                        \\ \cline{1-5}
14                        & 0.0305                                                        & 11                        & 0.066516603                                & CFR12              &                                                           &                        \\ \cline{1-5}
9                         & 0.0325                                                        & 6                         & 0.070802932                                & CFR13              &                                                           &                        \\ \cline{1-5}
10                        & 0.0321                                                        & 7                         & 0.069852078                                & CFR14              &                                                           &                        \\ \cline{1-5}
                          & 0.4592                                                        &                           & 1                                          & \textbf{SUM}       & \multirow{-15}{*}{0.45918}                                & \multirow{-15}{*}{CFR} \\ \hline
                          &                                                               &                           & 1                                          & \textbf{SUM}       &                                                           &                        \\ \hline
                          & 1                                                             & \textbf{SUM}              &                                            &                    &                                                           &                        \\ \hline
\end{tabular}
\caption{Average weights of the main and secondary criteria of the five experts.}~\label{avertab001}
\end{table}
\end{itemize}
The previous table shows the average weights of primary and secondary standards according to the five expert opinions, where it was noted that 
The risks derived from the cash flow statement are the most serious in terms of the degree of importance relative to expert opinions, followed by liquidity risks, then risks derived from the Income Statement, and the least serious are the capital structure risks, which confirms the importance of risks derived from the cash flow statement, as this list provides information that other financial statements do not provide, such as the company's ability to generate cash from its operating, investment and financing activities sufficient to meet its obligations~\cite{Yasmin2016}, and the extent of its need for external financing~\cite{Nubani2011}, any assessment of the company's ability to generate cash from its operating, investment and financing activities sufficient to meet its obligations~\cite{Yasmin2016}, and the extent of its need for external financing~\cite{Nubani2011} the financial situation of the company is better.
Among Figure~\ref{figmarwasdn1} the average weights of the main criteria for the goal shown in the previous Table~\ref{avertab001}, we find:
\begin{figure}[H]
    \centering
    \includegraphics[width=0.5\linewidth]{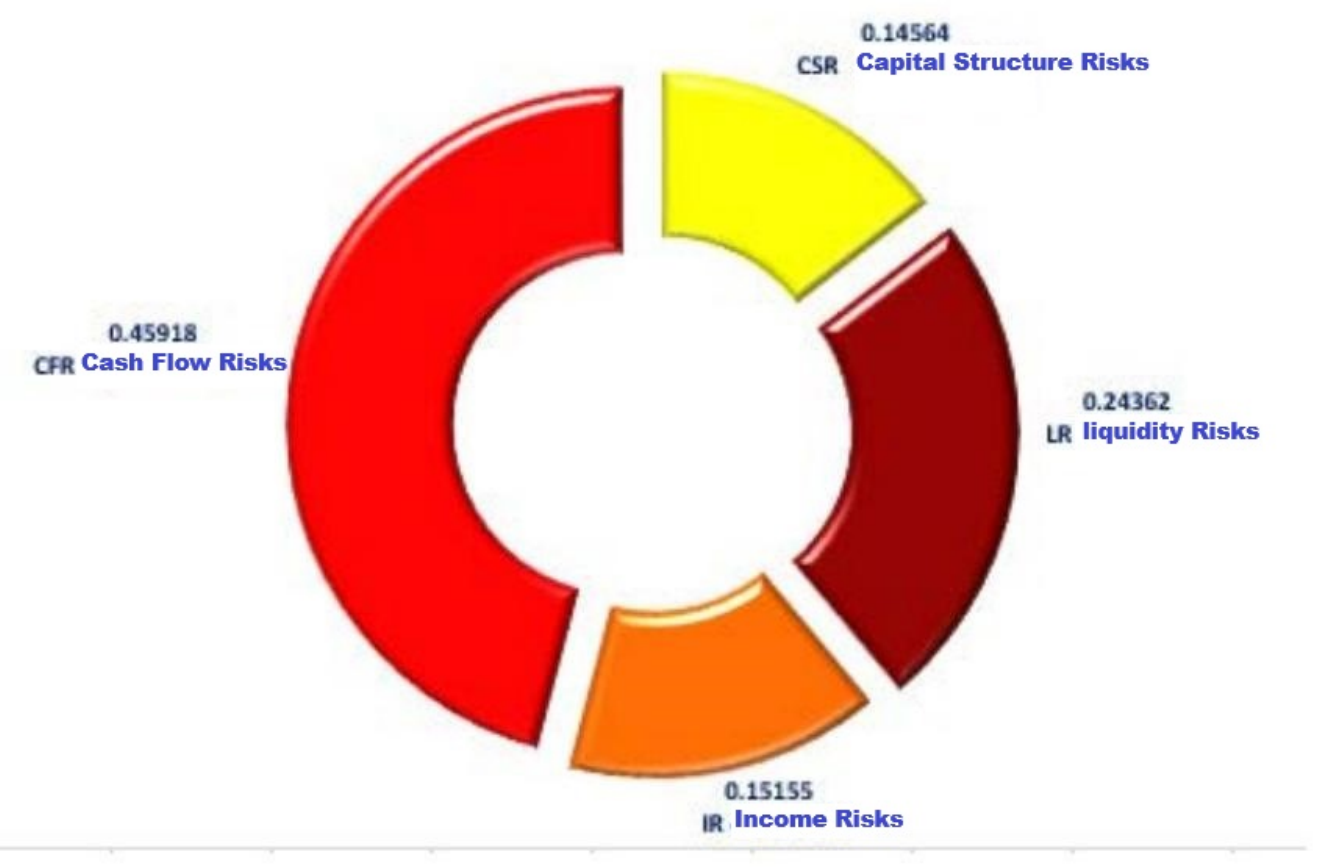}
    \caption{Average weights of the main criteria.}
    \label{figmarwasdn1}
\end{figure}
It is noted that the risks derived from the cash flow statement received the highest relative weight at 45.9\%, and the capital structure risks received the lowest relative weight at 14.5\%. 
As for the graphical representation of the average weights of secondary standards relative to the target (see Figure~\ref{figmarwasdn2}), which are also shown in Table~\ref{avertab001}, it was as follows:
\begin{figure}[H]
    \centering
    \includegraphics[width=0.8\linewidth]{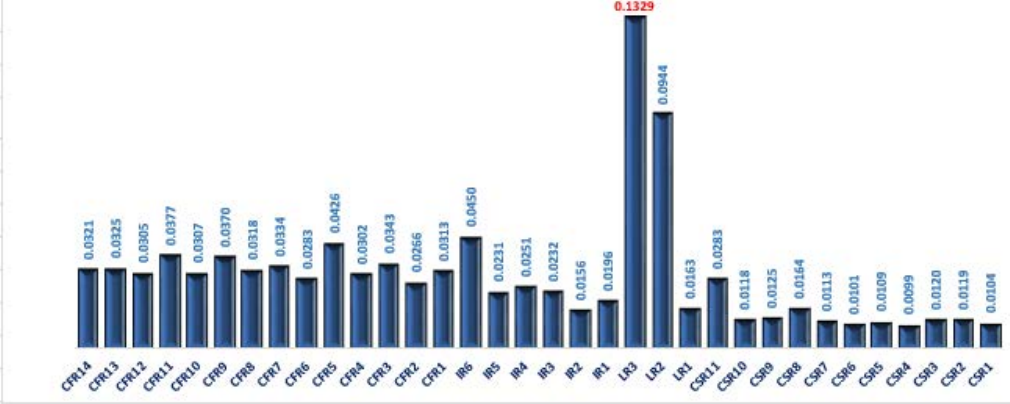}
    \caption{Average expert opinions.}
    \label{figmarwasdn2}
\end{figure}
It is noticeable that the liquidity readiness ratio (LR3) is the most critical, having reached 13.2\%, as it expresses the company's readiness to fulfill its short-term and immediate obligations~\cite{Din2016}, and that the retained earnings to assets ratio (CSR4) is the least risky, reaching 90.9\%. This ratio represents an internally generated funding source and reflects a temporary stock of funds within the company to deal with losses~\cite{Btach2010}.

From the foregoing, it is clear that we used AHP up to the third level of the hierarchical model (Figure~\ref{figmarwasdn2}), which is the secondary criteria. To complement AHP with SAW, we apply the previous SAW steps to evaluate the alternatives (financial years), which represent the base of the pyramid in the model.

\begin{enumerate}
    \item \textbf{Integration between AHP and SAW:}
    \begin{enumerate}
        \item Building the initial decision matrix: The matrix consists of ten rows representing the alternatives, which are the financial years, and thirty-four columns representing the financial ratios mentioned in Tables~\ref{tabmarn1},\ref{tab005Standardsr} and \ref{avertab001}, where the intersection of a row and column represents the value of the criterion (financial ratio) for each alternative (financial year).
Author's calculated the financial ratios mentioned in Tables~\ref{tabmarn1},\ref{tab005Standardsr} and \ref{avertab001} for the National Vegetable Oils Company for a period of ten years from 2008 to 2017 and used them as columns in the initial decision matrix.
\item \textbf{Normalization of the Initial Decision Matrix using the Max-Min Method}

To normalize the initial decision matrix, we must first identify the benefit criterion (Max) and the cost criterion (Min) for each column (financial ratio). This determination is based on what leads to the reduction of financial risk.

For instance, we presented among Table~\ref{tab001matrix} the ratio of net operating cash flow to total debt (CFR11) is considered a benefit criterion (Max), because as its value increases, the associated financial risk decreases. Conversely, the ratio of total debt to total equity (CSR1) is considered a cost criterion (Min), as an increase in its value signifies a higher financial risk. This can be summarized as shown in the following table:
\begin{table}[H]
    \centering
    \begin{tabular}{|l|c||l|c|}
    \hline
  CSR1  & Min & LR1   & Max \\ \hline
CSR2  & Min & LR2   & Max \\ \hline
CSR3  & Min  & LR3   & Max \\ \hline
CSR4  & Max & IR1   & Min \\ \hline
CSR5  & Min & IR2   & Max \\ \hline
CSR6  & Min & IR3   & Max \\ \hline
CSR7  & Min & IR4   & Max \\ \hline
CSR8  & Max & IR5   & Min \\ \hline
CSR9  & Max & IR6   & Min \\ \hline
CSR10 & Max & CSR11 & Max \\ \hline
CFR1  & Max & CFR2  & Max \\ \hline
CFR3  & Max & CFR4  & Max \\ \hline
CFR5  & Max & CFR14 & Max \\ \hline
CFR6  & Max& CFR7  & Max \\ \hline
CFR8  & Max & CFR9  & Max \\ \hline
CFR10 & Max & CFR11 & Max \\ \hline
CFR12 & Max & CFR13 & Max \\ \hline
\end{tabular}
\caption{Determination of the utility or cost criterion for each column in the initial decision matrix.}
\label{tab001matrix}
\end{table}
Next, we apply the aforementioned normalization method (Step 2 of the SAW method) to the entire initial decision matrix. As a result, the values of the elements in this matrix are transformed into positive values falling within the range $[0, 1]$.
% --- ----------------------------------------
\item \textbf{Constructing the Weighted Decision Matrix}

The values of this matrix are calculated by applying the following relation to the elements of the normalized matrix:
\begin{equation} \label{eq:weighted_matrix}
v_{ij} = w_j \times r_{ij}
\end{equation}

Where:
\begin{itemize}
    \item[\(w_j\):] Represents the criteria weights obtained from the AHP method. These weights are the average of the opinions of the five experts at the National Company for Vegetable Oils, as shown in Table (9).
    \item[\(r_{ij}\):] Represents the values of the elements in the normalized matrix.
\end{itemize}

After this step, we obtain the weighted decision matrix, which consists of 10 rows (financial years) and 34 columns (financial ratios), Table as follows:

\begin{sidewaystable}[!htbp]
    \centering
    \small
    \begin{tabular}{|l|c|c|c|c|c|c|c|c|c|c|}
        \hline
        \cellcolor{green}\textbf{Year} & \textbf{CSR1} & \textbf{CSR2} & \textbf{CSR3} & \textbf{CSR4} & \textbf{CSR5} & \textbf{CSR6} & \textbf{CSR7} & \textbf{CSR8} & \textbf{CSR9} & \textbf{CSR10} \\
        \hline
        \textbf{2008} & 0         & 0.011618 & 0.011538 & 0.000323 & 0.010413 & 0         & 0.009664 & 0         & 0.0003   & 0.011834 \\ \hline
        \textbf{2009} & 0.0057433 & 0.007244 & 0.006102 & 0        & 0.005172 & 0.0055680 & 0        & 0.000195  & 0        & 0.005282 \\ \hline
        \textbf{2010} & 0.0103733 & 0.011905 & 0        & 0.000629 & 0        & 0.0100567 & 0.003473 & 0.000475  & 0.000197 & 0        \\ \hline
        \textbf{2011} & 0.0059623 & 0.006533 & 0.012021 & 0.002116 & 0.010919 & 0.0057803 & 0.011344 & 0.001014  & 0.000592 & 0.005032 \\ \hline
        \textbf{2012} & 0.0074227 & 0.008156 & 0.012021 & 0.006647 & 0.010919 & 0.0071961 & 0.011344 & 0.002218  & 0.001523 & 0.003366 \\ \hline
        \textbf{2013} & 0.0037084 & 0.004028 & 0.012021 & 0.00987  & 0.010919 & 0.0035952 & 0.011344 & 0.003787  & 0.002738 & 0.007603 \\ \hline
        \textbf{2014} & 0.0009759 & 0.000992 & 0.012021 & 0.003155 & 0.010919 & 0.0009462 & 0.011344 & 0.005189  & 0.003823 & 0.010721 \\ \hline
        \textbf{2015} & 0.0006741 & 0.000657 & 0.012021 & 0.001889 & 0.010919 & 0.0006535 & 0.011344 & 0.006192  & 0.006609 & 0.011065 \\ \hline
        \textbf{2016} & 0.0057265 & 0.006271 & 0.012021 & 0.003737 & 0.010919 & 0.0055517 & 0.011344 & 0.016319  & 0.012436 & 0.005301 \\ \hline
        \textbf{2017} & 0.0000832 & 0        & 0.012021 & 0.003066 & 0.010919 & 0.0000806 & 0.011344 & 0.016445  & 0.012534 & 0.011739 \\ \hline
        \cellcolor{green}\textbf{Year} & \textbf{CSR11} & \textbf{LR1} & \textbf{LR2} & \textbf{LR3} & \textbf{IR1} & \textbf{IR2} & \textbf{IR3} & \textbf{IR4} & \textbf{IR5} & \textbf{IR6} \\
        \hline
        \textbf{2008} & 0.007802 & 0.016295 & 0.053361 & 0.0237    & 0.014338 & 0        & 0.002375 & 0.003128 & 0.019686 & 0.036792 \\ \hline
        \textbf{2009} & 0        & 0.000848 & 0.067017 & 0         & 0.017351 & 0.000299 & 0.001948 & 0.002288 & 0.019509 & 0.037756 \\ \hline
        \textbf{2010} & 0.001883 & 0.015239 & 0.094438 & 0.089729  & 0.018714 & 0.002686 & 0.002778 & 0.003248 & 0.01828  & 0.036064 \\ \hline
        \textbf{2011} & 0.005212 & 0.001258 & 0.064656 & 0.013633  & 0.019545 & 0.0041   & 0.003916 & 0.004575 & 0.016158 & 0.031718 \\ \hline
        \textbf{2012} & 0.011197 & 0.004515 & 0.08144  & 0.095478  & 0.019123 & 0.009588 & 0.007096 & 0.008137 & 0.009737 & 0.019172 \\ \hline
        \textbf{2013} & 0.016193 & 0.001185 & 0.069723 & 0.051551  & 0.019114 & 0.015581 & 0.013744 & 0.015813 & 0.008973 & 0.016329 \\ \hline
        \textbf{2014} & 0.019201 & 0        & 0.068099 & 0.044844  & 0.018431 & 0.009212 & 0.010004 & 0.01146  & 0.009456 & 0.015606 \\ \hline
        \textbf{2015} & 0.023885 & 0.000369 & 0.072524 & 0.063616  & 0.018809 & 0.005995 & 0.013528 & 0.014968 & 0.004364 & 0.004911 \\ \hline
        \textbf{2016} & 0.028294 & 0.005288 & 0.089033 & 0.13289   & 0.01962  & 0.013289 & 0.023157 & 0.02514  & 0        & 0        \\ \hline
        \textbf{2017} & 0.007802 & 0.000558 & 0.058713 & 0.002747  & 0        & 0.002068 & 0        & 0        & 0.023101 & 0.044955 \\ \hline
         \cellcolor{green}\textbf{Year} & \textbf{CFR1} & \textbf{CFR2} & \textbf{CFR3} & \textbf{CFR4} & \textbf{CFR5} & \textbf{CFR6} & \textbf{CFR7} & \textbf{CFR8} & \textbf{CFR9} & \textbf{CFR10} \\
        \hline
        \textbf{2008} & 0.019367 & 0.016741 & 0.020813 & 0.013716 & 0.001568 & 0.016813 & 0.020708 & 0.000255 & 0.034313 & 0.023289 \\ \hline
        \textbf{2009} & 0.018719 & 0.017762 & 0.020355 & 0.013034 & 0.002815 & 0.018178 & 0.022177 & 0.006967 & 0.034059 & 0.024037 \\ \hline
        \textbf{2010} & 0.015798 & 0.021209 & 0.019742 & 0.030224 & 0.004235 & 0.021455 & 0.025507 & 0.031805 & 0.029426 & 0.02561  \\ \hline
        \textbf{2011} & 0.031334 & 0.013811 & 0.022439 & 0.007964 & 0        & 0.013843 & 0.017235 & 0        & 0.034568 & 0.021667 \\ \hline
        \textbf{2012} & 0        & 0.026639 & 0        & 0.027406 & 0.003144 & 0.028341 & 0.033448 & 0        & 0        & 0.030666 \\ \hline
        \textbf{2013} & 0.017983 & 0.01995  & 0.014707 & 0.012358 & 0.001253 & 0.018391 & 0.022503 & 0        & 0.031205 & 0.024736 \\ \hline
        \textbf{2014} & 0.017496 & 0.021703 & 0.016594 & 0.01358  & 0.001691 & 0.02083  & 0.025583 & 0        & 0.028039 & 0.027063 \\ \hline
        \textbf{2015} & 0.016514 & 0.020155 & 0.019816 & 0.012663 & 0.001282 & 0.019566 & 0.024063 & 0        & 0.029753 & 0.026922 \\ \hline
        \textbf{2016} & 0.016564 & 0.023707 & 0.017515 & 0.015175 & 0.0013   & 0.020506 & 0.024837 & 0        & 0.026924 & 0.029753 \\ \hline
        \textbf{2017} & 0.025853 & 0        & 0.034289 & 0        & 0.042565 & 0        & 0        & 0        & 0.03703  & 0        \\ \hline
    \end{tabular}
\end{sidewaystable}
\begin{table}[H]
    \centering
    \begin{tabular}{|l|c|c|c|c|}
        \hline
        \cellcolor{green}\textbf{Year} & \textbf{CFR11} & \textbf{CFR12} & \textbf{CFR13} & \textbf{CFR14} \\
        \hline
        \textbf{2008} & 0.014853 & 0.018144 & 0.030027 & 0.007045 \\ \hline
        \textbf{2009} & 0.017454 & 0.019039 & 0.029174 & 0.007413 \\ \hline
        \textbf{2010} & 0.036731 & 0.019039 & 0.028513 & 0.003293 \\ \hline
        \textbf{2011} & 0.010958 & 0.019039 & 0.032511 & 0.006224 \\ \hline
        \textbf{2012} & 0.037712 & 0        & 0        & 0        \\ \hline
        \textbf{2013} & 0.017005 & 0.015868 & 0.021218 & 0.004702 \\ \hline
        \textbf{2014} & 0.018686 & 0.015014 & 0.023941 & 0.004639 \\ \hline
        \textbf{2015} & 0.017425 & 0.010714 & 0.02859  & 0.002946 \\ \hline
        \textbf{2016} & 0.020881 & 0.002218 & 0.025269 & 0.004992 \\ \hline
        \textbf{2017} & 0        & 0.030543 & 0.026177 & 0.032075 \\ \hline
    \end{tabular}
    \caption{Complete Weighted Decision Matrix.}
    \label{tab0090complrte.}
\end{table}

\item Calculate the sum of each alternative (fiscal year) in the decision matrix burdened with the weights of the criteria by applying the following relationship:
\[
V = \sum_{j=1}^{m} w_jr_{ij}.
\]

\item Rank the alternatives:

\[
A_{ij}=\dfrac{V_{ij}}{\sum_{j=1}^{m}V_{ij}}.
\]

\item Determine the best alternative as in Table~\ref{tab009Weights} below: 
\begin{table}[H]
\centering
\begin{tabular}{|l|l|l|l}
\hline
Replacement  & V           & A     & \multicolumn{1}{l|}{Rank}                       \\ \hline
2008         & 0.47081576  & 0.097 & \multicolumn{1}{l|}{6}                          \\ \hline
2009         & 0.388439103 & 0.080 & \multicolumn{1}{l|}{\cellcolor{green}{10}} \\ \hline
2010         & 0.632756443 & 0.130 & \multicolumn{1}{l|}{2}                          \\ \hline
2011         & 0.408729192 & 0.084 & \multicolumn{1}{l|}{8}                          \\ \hline
2012         & 0.502288945 & 0.103 & \multicolumn{1}{l|}{3}                          \\ \hline
2013         & 0.479082381 & 0.098 & \multicolumn{1}{l|}{5}                          \\ \hline
2014         & 0.45797066  & 0.094 & \multicolumn{1}{l|}{7}                          \\ \hline
2015         & 0.4793678   & 0.098 & \multicolumn{1}{l|}{4}                          \\ \hline
2016         & 0.647092667 & 0.133 & \multicolumn{1}{l|}{\cellcolor{red}{1}}  \\ \hline
2017         & 0.408532659 & 0.084 & \multicolumn{1}{l|}{9}                          \\ \hline
\textbf{SUM} & 4.87507561  & 1.000 &                                                 \\ \cline{1-3}
\end{tabular}
\caption{Weights and order of substitutions.}~\label{tab009Weights}
\end{table}
    \end{enumerate}
\end{enumerate}
And by Figure~\ref{fig990alternatives} representation of the previous table we find:
\begin{figure}[H]
    \centering
    \includegraphics[width=0.5\linewidth]{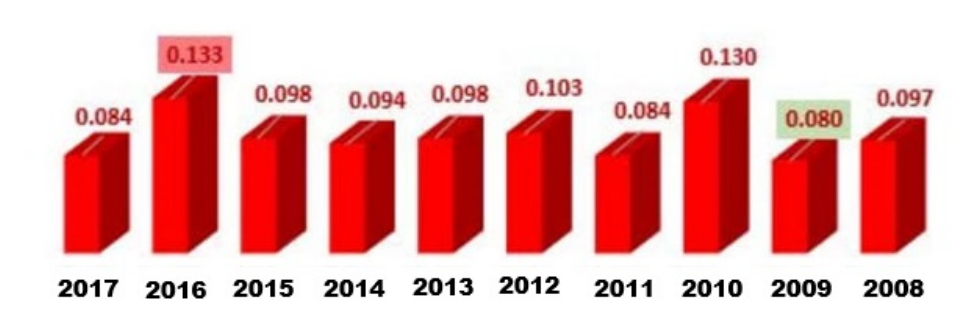}
    \caption{Weights of fiscal years (alternatives).}
    \label{fig990alternatives}
\end{figure}
We note from the previous figure that the most dangerous fiscal year is 2016.
%====================
\section{Conclusion}\label{sec5}
%====================
Through this paper, we established that 
\begin{enumerate}
    \item \textbf{the result of the AHP application shows the following:}
    \begin{enumerate}
        \item The risks derived from the cash flow statement are the most dangerous for the company, as it amounted to 45.9\%, reflecting its important role in the statement of financial risks facing the company.one of the most important ratios is the ratio of net operating cash flows to net profit, which is one of the important financial ratios that help the company in predicting the risks of bankruptcy and financial failure.
        \item The relative importance of liquidity risks has reached 24.3\% compared to other risks, and the most important percentage is the cash readiness ratio, which reflects the company's ability to face the risks of paying its short-term obligations when due without the need to borrow or liquidate one of its fixed assets.
        \item The risks derived from the income statement have a higher importance than the risks of the capital structure, where the first reached 15.1\%, and the second reached 14.5\%, which reflects the ability of income statement analysis to provide information on financial risks more important than the information provided by the analysis of its financial structure.
    \end{enumerate}
\item \textbf{show the following AHP and saw integration:}
The year 2016 was the year in which the company was exposed to the highest level of financial risk during the studied period, where its relative importance (Table No. 12) was 13.3\%, while 2009 is the least risky year, where its relative importance was 8\%. And that the four highest-risk financial ratios are HILR3, LR2, IR6 ,CFR5 (Table~\ref{avertab001}) and comparing the actual values of these ratios, which were calculated based on the company's financial statements for the years 2016 , 2009, we find:
\begin{table}[H]
    \centering
    \begin{tabular}{|l|c|c|}
    \hline
The ratio & 2016 & 2009 \\ \hline
Cash readiness ratio LR3 & 729\% & 45.2\% \\ \hline
Fast liquidity ratio LR2 & 759\% & 268\% \\ \hline
Net Profit / equity ratio IR6 & 37\% & 4.6\% \\ \hline 
Net cash flows from operating activities / net profit CFR5 & 44\% & 181\% \\ \hline
\end{tabular}
\end{table}
We note that the company is able to repay its short-term obligations by achieving high liquidity in 2016, while that capacity decreases significantly in 2009. And that it achieves a return from investing its own funds in 2016, higher than it achieved in 2009, that is, its risk in 2016 is higher than the risk in 2009 due to the close and well-known relationship between risk and return, the higher the return, the higher the risk, and the company's activity shows that it generated cash at a much higher rate in 2009 than in 2016, exposing the company to the highest level of financial risk during the period studied.
\item the integration between AHP and saw helped to determine the relative importance of the risks to which the company is exposed, so the company can identify and then evaluate its risks in order to avoid them in the future.
\end{enumerate}

\section{Recommendations}
\begin{enumerate}
    \item The company should be interested in managing financial risks, especially liquidity risks, to determine the extent of its ability to achieve its goals and its continuity.
\item The company should study and analyze its cash flows more accurately in order not to be exposed to the possibilities of hardship or financial failure.
\item The company should measure the financial risks, which enables it to disclose them.
\item The need for the company to determine the sources of financing carefully and in a timely manner.
\item The company should use multi-criteria decision-making techniques in solving problems of a financial nature, as they help to simplify those problems, making it easier to solve them.
\end{enumerate}

%===========================
\section*{Declarations}
\begin{itemize}
	\item Funding: Not Funding.
	\item Conflict of interest/Competing interests: The author declare that there are no conflicts of interest or competing interests related to this study.
	\item Ethics approval and consent to participate: The author contributed equally to this work.
	\item Data availability statement: All data is included within the manuscript.
\end{itemize}

\end{document}